\documentclass[twocolumn]{aastex}
\usepackage{amsmath,amsfonts, amssymb, url, natbib, enumerate, graphicx, mathtools,epstopdf}
\usepackage[normalem]{ulem}
\usepackage{color,url}

\defcitealias{Li2016}{L16}
\defcitealias{Keating2015}{K15}
\begin{document}
\submitted{Accepted for publication in ApJ}
\title{COPSS II: The molecular gas content of \\ ten million cubic megaparsecs at redshift \lowercase{$z\sim3$}}
\author{Garrett K. Keating\altaffilmark{1}, Daniel P. Marrone\altaffilmark{2}, Geoffrey C. Bower\altaffilmark{3}, \\ Erik~Leitch\altaffilmark{4,5}, John E. Carlstrom\altaffilmark{4} and David R. DeBoer\altaffilmark{1}}

\altaffiltext{1}{Radio Astronomy Laboratory, 501 Campbell Hall, University of California, Berkeley, CA 94720, USA; karto@astro.berkeley.edu}
\altaffiltext{2}{Steward Observatory, University of Arizona, 933 North Cherry Avenue, Tucson, AZ 85721, USA}
\altaffiltext{3}{Academia Sinica Institute of Astronomy and Astrophysics, 645 N. A'ohoku Pl., Hilo, HI 96720, USA; gbower@asiaa.sinica.edu.tw}
\altaffiltext{4}{Department of Astronomy \& Astrophysics, Kavli Institute for Cosmological Physics, University 
of Chicago, Chicago, IL 60637, USA}
\altaffiltext{5}{California Institute of Technology, Owens Valley Radio Observatory, Big Pine, CA 93513, USA}

\shorttitle{COPSS II}
\shortauthors{Keating et al.}
\begin{abstract}
We present a measurement of the abundance of carbon monoxide in the early Universe, utilizing the final results from the CO Power Spectrum Survey (COPSS). Between 2013 and 2015, we performed observations with the Sunyaev-Zel'dovich Array to measure aggregate CO emission from $z\sim3$ galaxies with the intensity mapping technique. Data were collected on 19 fields, covering an area of 0.7 square degrees, over the frequency range $27{-}35$ GHz. With these data, along with data analyzed in COPSS I, we are able to observe the CO(1-0) transition within the redshift range $z=2.3{-}3.3$ for spatial frequencies between $k=0.5{-}10\,\,h\,\textrm{Mpc}^{-1}$, spanning a comoving volume of $4.9\times10^{6}\,\,h^{-3}\,\textrm{Mpc}^{3}$. We present estimates of contributions from continuum sources and ground illumination within our measurement. We constrain the amplitude of the CO power spectrum to $P_{\textrm{CO}}=3.0_{-1.3}^{+1.3}\times10^{3}\,\,\mu\textrm{K}^{2}(h^{-1}\,\textrm{Mpc})^{3}$, or $\Delta^{2}_{\textrm{CO}}(k\!=\!1\,\,h\,\textrm{Mpc}^{-1})=1.5^{+0.7}_{-0.7}\times10^{3}\,\,\mu\textrm{K}^{2}$, at $68\%$ confidence, and $P_{\textrm{CO}}>0$ at 98.9\% confidence. These results are a factor of 10 improvement in sensitivity compared to those of COPSS I. With this measurement, we constrain on the CO(1-0) galaxy luminosity function at $z\sim3$. Assuming that CO emission is proportional to halo mass, and using theoretical estimates of the scatter in this relationship, we constrain the ratio of $\textrm{CO}(1{-}0)$ luminosity to halo mass to $A_{\textrm{CO}}=6.3_{-2.1}^{+1.4}\times10^{-7}\,\,L_{\odot}\,\,M_{\odot}^{-1}$. Assuming a Milky Way-like linear relationship between CO luminosity and molecular gas mass, we estimate a mass fraction of molecular gas of $f_{\textrm{H}_{2}}=5.5^{+3.4}_{-2.2}\times10^{-2}$ for halos with masses of $\sim10^{12}M_{\odot}$. Using theoretical estimates for the scaling of molecular gas mass fraction and halo mass,  we estimate the cosmic molecular gas density to be $\rho_{z\sim3}(\textrm{H}_{2})=1.1_{-0.4}^{+0.7}\times10^{8}\,\,M_{\odot}\,\,\textrm{Mpc}^{-3}$.
\end{abstract}
%%%%%%%%%%%%%%%%%%%%%%%%%%%%%%%%%%%%%%%%%%%%%%%%%%%
%%%%%% Introduction%%%%%%
\section{Introduction}\label{sec_intro}
The gas content of galaxies play a powerful role in shaping their evolution. In the early Universe, large gas reservoirs dominated the baryonic mass of galaxies \citep{Tacconi2010,Popping2015}, and fueled a rapid increase in cosmic star formation, peaking at a rate 10 times higher than what is observed locally \citep{Madau2014,Hopkins2006}. Most of our knowledge of these early galaxies has come from studying stellar light and emission lines from the hot, ionized gas of the interstellar medium (ISM). However, it is the cold molecular gas that provides the natal material from which stars form. Understanding the nature and evolution of this cold gas is thus crucially important for understanding star formation in the early Universe.

Of the several tracers available for studying cold gas in the local Universe, the bright transitions of carbon monoxide (CO) are particularly suitable for examining distant, high-redshift objects. The CO molecule is typically found in clouds of molecular hydrogen (for a review, see \citealt{Bolatto2013}). Sensitive instruments such as the VLA, ALMA and PdBI have made it possible to probe the cool ISM of massive galaxies out to redshifts as high as $z=6.4$ (e.g., \citealt{Walter2003,Wang2010,Riechers2013,Lentati2015}). However, these extraordinary objects are not likely to be characteristic of the overall population of star-forming galaxies in the early Universe, made up primarily of smaller and less luminous systems \citep{Bouwens2012,Smit2012,Robertson2015}.

Unfortunately, the molecular line emission from ``normal'' (i.e., low-mass) star-forming galaxies at $z\gtrsim3$ is extremely faint. While measurements of such objects are vital for characterizing normal star-forming galaxies, direct detection of the CO emission arising from individual high-redshift low-mass galaxies is observationally expensive, and the detection of significant numbers of such objects is likely out of the reach of the current generation of radio instruments \citep{Carilli2013}. In part, this has left us with an incomplete understanding of the cool ISM in early galaxies, and is a key limitation in using galaxy formation simulations to understand even the most luminous of galaxies (e.g., \citealt{Hayward2011}). 

One alternative method to exploring the ISM within more typical galaxies in the early Universe is through a technique commonly referred to as ``intensity mapping'', where emission from a multitude of galaxies (i.e., thousands or millions) over a wide range of luminosities is detected in aggregate as fluctuations in the mean line intensity over large spatial scales. The intensity mapping method is a valuable tool for charting the growth of large scale structure and of the gas contents of galaxies. This method has been the subject of numerous recent theoretical investigations (e.g., \citealt{Righi2008,Visbal2010,Carilli2011,Pullen2013,Breysse2014,Mashian2015,Li2016}), several of which suggest that such a signal may be detected with existing instruments. Intensity mapping experiments are well-suited for data sets with large survey volumes, requiring only modest point-source sensitivity to detect an aggregate signal. As such, the Sunyaev-Zel'dovich Array (SZA) -- a 3.5m-diameter $\times$ 8-element subset of the Combined Array for Research in Millimeter-wave Astronomy (CARMA) -- is an instrument suited to such an experiment.

The research presented here is the second phase of the CO Power Spectrum Survey (COPSS) -- an experiment designed to detect the aggregate CO signal of the early universe. The first phase of this experiment (COPSS I; discussed in \citealt{Keating2015}, hereafter referred to as \citetalias{Keating2015}) made use of archival SZA data to place the first constraints on the CO autocorrelation power spectrum at $z\sim3$. In this paper we present results from an observing campaign dedicated to deep integrations that could detect the CO power spectrum signal. Using three times the integration time, a compact array configuration that nearly doubles our sensitivity, and deeper integration on a smaller number of fields, we achieve greatly improved sensitivity. We have structured this paper in the following way: We  discuss the COPSS survey and the SZA instrument in Section~\ref{sec_data}. In Section~\ref{sec_analysis}, we discuss the analysis procedures and tests for systematic errors. Section~\ref{sec_results} describes the results of the survey, followed by their implications in in Section~\ref{sec_discussion}. Conclusions are presented in Section~\ref{sec_conclusion}. Throughout this paper, we assume a $\Lambda$CDM cosmology, with $h=0.7$, $\Omega_{\textrm{m}}=0.27$, $\Omega_{\Lambda}=0.73$.

%%%%%%%%%%%%%%%%%%%%%%%%%%%%%%%%%%%%%%%%%%%%%%%%%%%
%%%%%% Observation%%%%%%
\section{Data}\label{sec_data}
%%%%%% SZA Description%%%%%%
\subsection{Instrument Description}\label{ssec_arraydescrip}
The Sunyaev-Zel'dovich Array is an 8-element subset of the Combined Array for Research in Millimeter-wave Astronomy. Each SZA antenna receives left-circular polarized light over a frequency range of 27$-$35~GHz. The 3.5m-diameter antennas provide a $\theta_{\textrm{B}} \approx 11'$ full width at half maximum primary beam at 31~GHz (solid angle $\Omega_{\textrm{B}}=3.8\times10^{-2}\ \textrm{deg}^{2}$), and have a typical aperture efficiency of 0.6. Under average weather conditions, the system temperature is $T_{\textrm{sys}}\approx 40\ \textrm{K}$. Except for the pilot observations (Table~\ref{table_fields}), which were taken with the array in a six-antenna compact configuration with two outriggers as in \citetalias{Keating2015}, the data presented here were obtained with all eight antennas in a compact configuration (4.5-16m spacings) to maximize sensitivity to large angular scales. Additional details about the SZA can be found in \cite{Muchovej2007} and \citetalias{Keating2015}.
%%%%%% COPSS II Description%%%%%%
\subsection{Survey Description}\label{ssec_observations}
\floattable
\begin{deluxetable*}{cccccc}
\tablewidth{500px}
\tablecaption{COPSS II Observing Fields.
\label{table_fields}}
\tablehead{\colhead{Field Name} & \colhead{RA} & \colhead{Dec} & \colhead{Gain Cal} & \colhead{Obs Time} & \colhead{Notes} \\
 & & & & (hours) & \colhead{Notes}}
\startdata
\hline
FLANK1-L  & 06$^{\textrm{h}}$33$^{\textrm{m}}$28$^{\textrm{s}}$.0 & +62$\degr$13$'$53$''$ & 3C147     & 221.6 &  a  \\
FLANK1    & 06$^{\textrm{h}}$38$^{\textrm{m}}$50$^{\textrm{s}}$.0 & +62$\degr$14$'$00$''$ & 3C147     & 270.3 & a,b \\
FLANK1-T  & 06$^{\textrm{h}}$44$^{\textrm{m}}$11$^{\textrm{s}}$.0 & +62$\degr$14$'$07$''$ & 3C147     & 226.8 &  a  \\
\hline
GOODS-NL  & 12$^{\textrm{h}}$31$^{\textrm{m}}$28$^{\textrm{s}}$.3 & +62$\degr$14$'$01$''$ & J1153+495 & 13.5  &  a  \\
GOODS-N   & 12$^{\textrm{h}}$36$^{\textrm{m}}$50$^{\textrm{s}}$.0 & +62$\degr$14$'$00$''$ & J1153+495 & 592.9 & a,b \\
GOODS-NT  & 12$^{\textrm{h}}$42$^{\textrm{m}}$11$^{\textrm{s}}$.8 & +62$\degr$13$'$59$''$ & J1153+495 & 280.7 &  a  \\
GOODS-T2  & 12$^{\textrm{h}}$47$^{\textrm{m}}$33$^{\textrm{s}}$.5 & +62$\degr$13$'$57$''$ & J1153+495 & 268.8 &  a  \\
\hline
AEGIS-L   & 14$^{\textrm{h}}$14$^{\textrm{m}}$09$^{\textrm{s}}$.6 & +52$\degr$51$'$04$''$ & J1419+543 &  95.8 &  a  \\
AEGIS     & 14$^{\textrm{h}}$19$^{\textrm{m}}$31$^{\textrm{s}}$.0 & +52$\degr$51$'$00$''$ & J1419+543 & 101.8 &  a  \\
AEGIS-T   & 14$^{\textrm{h}}$24$^{\textrm{m}}$52$^{\textrm{s}}$.4 & +52$\degr$50$'$56$''$ & J1419+543 &  92.3 &  a  \\
\hline 
Q2343-L   & 23$^{\textrm{h}}$40$^{\textrm{m}}$44$^{\textrm{s}}$.2 & +12$\degr$49$'$13$''$ & 3C454.3   & 146.4 &  a  \\
Q2343     & 23$^{\textrm{h}}$46$^{\textrm{m}}$05$^{\textrm{s}}$.0 & +12$\degr$49$'$12$''$ & 3C454.3   & 167.4 &  a  \\
Q2343-T   & 23$^{\textrm{h}}$51$^{\textrm{m}}$25$^{\textrm{s}}$.8 & +12$\degr$49$'$12$''$ & 3C454.3   & 158.6 &  a  \\
\hline
SXDS-L    & 02$^{\textrm{h}}$17$^{\textrm{m}}$12$^{\textrm{s}}$.0 & -04$\degr$59$'$59$''$ & J0224+069 &  27.5 &  a  \\
SXDS      & 02$^{\textrm{h}}$18$^{\textrm{m}}$00$^{\textrm{s}}$.0 & -05$\degr$00$'$00$''$ & J0224+069 &  11.5 &  a  \\
SXDS-T    & 02$^{\textrm{h}}$18$^{\textrm{m}}$48$^{\textrm{s}}$.0 & -05$\degr$00$'$01$''$ & J0224+069 &  21.6 &  a  \\
\hline
FLANK2    & 09$^{\textrm{h}}$33$^{\textrm{m}}$00$^{\textrm{s}}$.0 & +62$\degr$14$'$00$''$ & J0841+708 & 145.2 &  b  \\
FLANK3    & 15$^{\textrm{h}}$39$^{\textrm{m}}$02$^{\textrm{s}}$.0 & +62$\degr$14$'$00$''$ & J1642+689 &  38.7 &  b  \\
FLANK4    & 18$^{\textrm{h}}$11$^{\textrm{m}}$02$^{\textrm{s}}$.0 & +62$\degr$14$'$00$''$ & 3C371     & 103.4 &  b \\
\enddata
\tablenotetext{a}{Observed during primary survey}
\tablenotetext{b}{Observed during pilot survey}
\end{deluxetable*}
Data were collected during two different phases: pilot observations and primary survey observations. Primary survey observations were conducted between October 2014 and April 2015, and consist of 19 telescope pointings arranged into 6 groups. Pilot observations were conducted between April 2013 and April 2014, and consist of 5 telescope pointings. Primary survey fields -- FLANK1, GOODS-N \citep{Dickinson2003}, AEGIS \citep{Davis2007}, Q2343\citep{Steidel2004}, and SXDS \citep{Furusawa2008} -- for both phases were selected to allow for continuous 24-hour observations. GOODS-N, AEGIS, Q2343 and SXDS were also selected based on present and future availability of optical spectroscopic data (e.g., \citealt{Reddy2006,Brammer2012,Steidel2014,Kriek2015}), to enable potential future cross-correlation experiments.

The observations were structured to provide the ability to search for and remove contamination that depends on the telescope orientation, particularly emission from the ground and antenna cross-talk. \citetalias{Keating2015} found that ground contamination was a significant foreground when not removed. The pilot observations of GOODS-N were accompanied by observations of other fields (FLANK1--4) at the same declination, ensuring that the data sample the same locations in the $uv$ plane. These fields were widely spaced in RA to fill gaps in the CARMA observing schedule, and were therefore taken at different times than the GOODS-N pilot data. In the primary survey the main target fields (i.e, Q2343, SXDS, FLANK1, GOODS-N, and AEGIS) were accompanied by ``leading'' and ``trailing'' fields, separated from the main field by roughly 5 minutes in RA, such that the series of three fields was observed over the same hour angle over a series of sequential 5 minute observations. The one exception was GOODS-N, where two trailing fields were observed due to the presence of a strong $\sim40$ mJy point source in the leading field that was discovered during the first week of observations. For both phases of observations, fields were observed for a total of 180 5-second integrations (spent on one field for pilot observations, split between three fields during the primary survey), after which a gain calibrator was observed for several minutes, for a total of 20 minutes per observing loop. A bandpass and flux calibrator were typically observed between groups of fields, for 10 and 5 minutes respectively.

Provided in Table~\ref{table_fields} is a listing of the position, observing and integration time for each field.
%%%%%%%%%%%%%%%%%%%%%%%%%%%%%%%%%%%%%%%%%%%%%%%%%%%
%%%%%% Analysis%%%%%%
\section{Analysis}\label{sec_analysis}
%%%%%% Pipeline Overview %%%%%%
\subsection{Pipeline Overview}\label{ssec_pipeline}
In this section we provide a brief overview of the analysis techniques and software used in our analysis; a more detailed description of this software can be found in \citetalias{Keating2015}. Details of the power spectrum analysis and null tests can be found in Sections~\ref{ssec_powspecgen} and~\ref{ssec_jackknife}, respectively. The routines described herein were developed using MATLAB\footnote{Mathworks, Version 2013b, \url{http://www.mathworks.com/products/matlab/}}. 

The calibration procedures for the SZA data are similar to that used in \citetalias{Keating2015}. Raw data from the interferometer are recorded as visibilities, and converted to physical units using system temperature measurements that are made once every 20 minutes (i.e, the length of a source-calibrator cycle). Absolute aperture efficiencies are derived from observations of Mars, using the brightness temperature model from \citet{Rudy1987}. We expect that absolute flux measurements are accurate to within 10\%. A bright point source (within $20^{\circ}$ of the target fields being observed) is observed once every 20 minutes to provide relative gain calibration, along with a strong point source observed once every 6-hours to provide bandpass calibration. Once calibration solutions are derived, data from target fields are used to measure the difference between the measured and expected noise to determine an antenna-frequency channel dependent system equivalent flux density (SEFD) correction. Bad data are flagged in various stages throughout the calibration process.

Gain solutions for the SZA are generally very stable across both frequency and time. Gain phase and amplitude and phase vary by $<20^\circ$ and 3\%, respectively, over the course of a typical 24-hour period. As phase solutions are found to slowly drift over the course of a track, they are linearly interpolated between calibrator observations, whereas gain amplitudes are stable enough to be averaged over the track. Large gain changes ($>30\degr$ in phase or $>10\%$ in amplitude) are indications of potential data problems and target data between discrepant calibrator observations are flagged and excluded from subsequent analysis. We find that bandpass solutions are also very stable, with solutions typically showing less than 1\% variability between days, and an RMS variability of 1.2\% over the course of the entire survey.

\begin{figure*}[t]
\begin{center}
\includegraphics[scale=0.5]{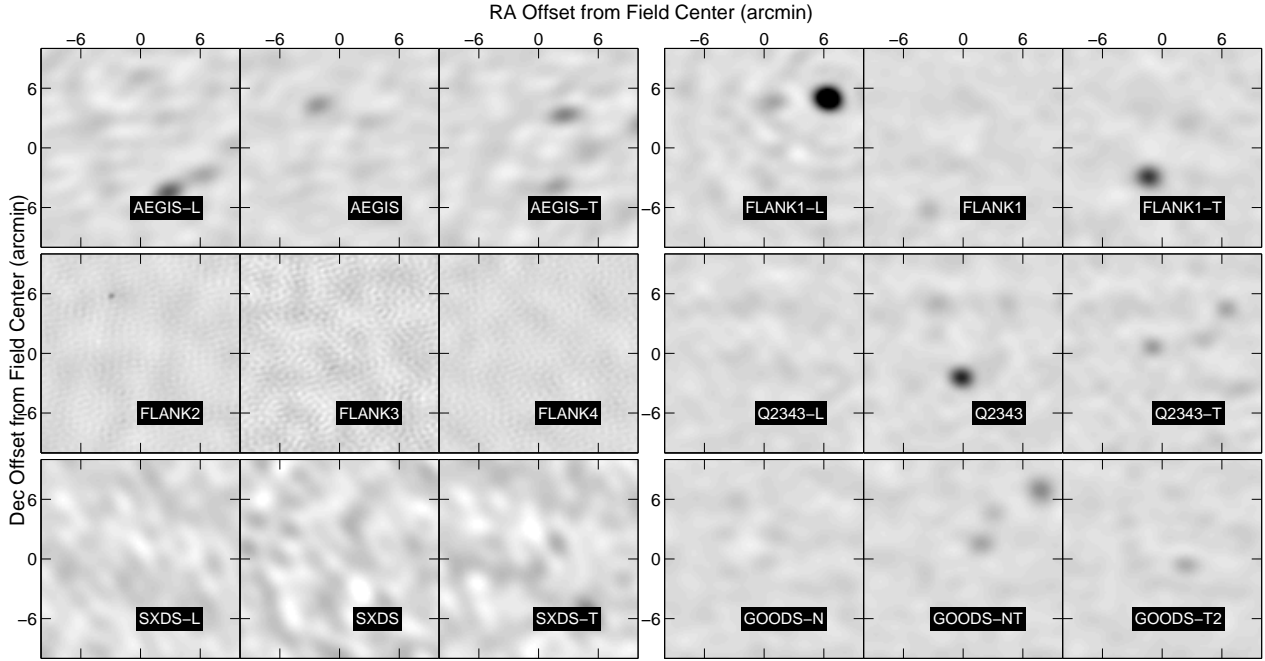}
\caption{Thumbnail images of the 18 target fields, after CLEAN deconvolution. An average of two continuum sources are detected per field above our $\sim$0.25~mJy threshold. The FLANK2-4 fields were observed entirely during the pilot phase and therefore show noise patterns modulated by the high-spatial-frequency data of the outrigger antennas. In other fields, these long baselines are absent or have too little sensitivity to be noticed in the maps.
\label{fig_thumbimages}}
\end{center}
\end{figure*}

Figure~\ref{fig_thumbimages} presents images (without primary beam correction) for all fields (with the exception of GOODS-NL). These images are deconvolved using the CLEAN algorithm \citep{Hogbom1974}. The typical synthesized beam for these images is $2'$. The median theoretical noise in the images is 0.05 mJy/beam, consistent with the RMS residual noise of the images after removal of detected point sources. Sources detected at more than 5$\sigma$ are fitted with point source models, and for sources above 10$\sigma$ we also allow spectral index freedom in the fits. These strong sources are subtracted from the visibilities to ensure that they do not introduce unexpected (real) correlation between visibilities that can be confused with false signals in subsequent flagging that relies on the statistical independence of the visibilities.

Once flagging, calibration, and point source removal are complete, ``delay-visibilities'' are produced by taking the Fourier transform of all visibilities within a single spectral window (within each baseline for a given integration). These delay-visibilities are then gridded in $(u,v,\eta,z)$ space, where $u$ and $v$ are the projected antenna spacings in the E-W and N-S coordinate directions, $\eta$ is the delay (i.e., Fourier dual of frequency), and $z$ is the median redshift of the spectral window.
%%%%%% Point Source %%%%%%
\subsection{Power Spectrum Analysis}\label{ssec_powspecgen}
For our analysis, we define
\begin{eqnarray}\label{eqn_powspec}
\Delta^{2}(k,z) &\equiv& \frac{k^{3}}{2\pi^{2}} P(k,z), \nonumber \\
&=& \frac{k^{3}}{2\pi^{2}} \frac{X^{4}Y^{2}}{\textrm{V}_{z}}\left(\frac{c^{2}}{2k_{B}\nu^{2}}\right)^{2} \left \langle \left | \tilde{I}^{2}\right | \right \rangle_{\mathbf{k}\cdot\mathbf{k}=k^{2}},
\end{eqnarray}
where the power spectrum, $P(k,z)$, is a measure of the variance in brightness temperature over a given comoving volume, expressed in this paper in units of $\mu\textrm{K}^{2}\ h^{-3}\,\textrm{Mpc}^{3}$ (where $h=H_{0}/100\ \textrm{km}/\textrm{s}/\textrm{Mpc}$; $H_{0}$ represents the current Hubble parameter). The power spectrum is given as a function of comoving spatial frequency $k$ (with units of $h$~Mpc$^{-1}$) and redshift $z$, and is proportional to the Fourier transform of the autocorrelation function of the intensity field under consideration (with $\tilde{I}$ defined as the Fourier transform of this intensity field). Given the solid angle of the telescope primary beam, $\Omega_{\mathrm{B}}$, and bandwidth, $B_{z}$, the volume surveyed by our measurement at a given redshift is $\textrm{V}_{z}=X^{2}YB_{z}\Omega_{\textrm{B}}/2$. $X$ and $Y$ are conversion factors between comoving distance and angular/frequency separation, respectively (e.g., \citealt{Parsons2012}). The Boltzmann constant is represented by $k_{B}$, the speed of light by $c$, and $\Delta^{2}(k)$ is the variance in brightness temperature per ln($k$), expressed in this paper in units of $\mu\textrm{K}^{2}$.

For the analysis presented here, we use two different methods to calculate power spectrum values: measuring the product of semi-correlated pairs (PSCP) of gridded delay-visibilities, and performing a maximum-likelihood evaluation (MLE) of the band-averaged power.  In \citetalias{Keating2015}, power spectrum values were calculated solely via the PSCP method using the following equation:
\begin{eqnarray}
\mathcal{P}(\mathbf{k},z) &=& \frac{\sum\limits_{\mathbf{k}^\prime}\sigma_{k}^{-2}\sigma_{k^{\prime}}^{-2} C(\mathbf{k}-\mathbf{k}^{\prime})\left( | \tilde{I}^{*}(\mathbf{k},z)\tilde{I}(\mathbf{k}^{\prime},z) |\right)}{\sum \limits_{\mathbf{k}^\prime} \sigma_{k}^{-2} \sigma_{k^{\prime}}^{-2} C^{2}(\mathbf{k}-\mathbf{k}^{\prime})} - \mathcal{A}_{\mathbf{k}} \nonumber \\
P(k,z) &=& \left \langle \mathcal{P}(\mathbf{k},z) \right \rangle_{\mathbf{k}\cdot\mathbf{k}=k^{2}} \label{eqn_powspec_pscp}.
\end{eqnarray}
In Equation~\ref{eqn_powspec_pscp}, $\mathcal{P}(\mathbf{k},z)$ is the three dimensional power spectrum, as a function of vector wavenumber, $\mathbf{k}$, and redshift. $\tilde{I}(\mathbf{k},z)$ represents the individual mode measurements (i.e., the renormalized delay-visibilities), $\sigma_{k}$ is the estimated thermal noise, and $C(\delta \mathbf{k})$ is the expected normalized covariance (for the signal of interest) for data separated by $\delta \mathbf{k}$ in the $(u,v,\eta)$ domain. In the PSCP method, all data within a single redshift window are cross-multiplied against one another, weighted by their estimated noise variance and signal covariance between cross-multiplied data. To remove noise bias, the sum of the autocorrelations of the individual delay-visibilities within each grid cell, $\mathcal{A}_{\mathbf{k}}$, is subtracted from our measurement. The PSCP method is computationally fast, requiring a few seconds of CPU time to calculate a power spectrum for an individual field. The primary limitation of this method is that it presumes the power spectrum errors are normally distributed (by way of the central limit theorem), when the distribution is actually a $\chi^{2}$ distribution with $N_{k}$ degrees of freedom (where $N_{k}$ is the number of independent measurements contained within each bin of the power spectrum). We note that the analysis within \citetalias{Keating2015} bore out the assumption of normally distributed errors for the COPSS I dataset, though those data provided many more independent measurements due to the increased number of fields (with near-equal sensitivity). 

Following the prescription from \citet{Bond1998} and \citet{Hobson2002}, under the assumption that our signal of interest comprises Gaussian fluctuations in brightness temperature, the likelihood of a given model for these fluctuations can be expressed as
\begin{equation}\label{eqn_powspec_mle}
\mathcal{L}(C) = \frac{1}{2\pi^{N/2}|C|^{1/2}}\textrm{exp}\left ( \frac{1}{2} \boldsymbol{\tilde{I}^{T}} C^{-1} \boldsymbol{\tilde{I}} \right)
\end{equation}
In Equation \ref{eqn_powspec_mle}, $\boldsymbol{\tilde{I}}$ is the vector containing the renormalized delay-visibilities, $C$ is the covariance matrix of the data set, and $N$ is the number of independent measurements within the data set. The covariance matrix can further be expressed as 
\begin{equation}\label{eqn_mle_constraints}
C = P(z) C_{\textrm{signal}} + C_{\textrm{noise}} + C_{\textrm{const}},
\end{equation}
Where $C_{\textrm{signal}}$ is the covariance induced by the signal of interest (calculated in the same fashion as the PSCP method, discussed in detail in \citetalias{Keating2015}), and $P(z)$ is the band-averaged power of the power spectrum at a given redshift (i.e., our simple model assumes that all modes measured within a single redshift window have power $P(z)$). $C_{\textrm{noise}}$ is a diagonal matrix containing the contribution of  instrumental noise to the measurement,  and $C_{\textrm{const}}$ is the ``constraint matrix'', used to downweight mode(s) with known unwanted contributions to the power spectrum (e.g., ground contaminants). The MLE method is computationally expensive, but it allows for the direct calculation of uncertainties in the power spectrum constraints, rather than relying on the central limit theorem approximations of the PSCP results.

In our analysis, we find that PSCP and MLE methods produce values that are generally within $0.1\sigma$ of each other, with errors that agree to within a few percent. Except where otherwise noted, we use the PSCP method to produce power spectrum figures, and otherwise use the MLE values in our analysis.
%%%%%% Ground Subtraction%%%%%%
\subsubsection{Ground Contamination and Subtraction}\label{ssec_groundsub}

Faint fluctuations from CO line emitters can easily be swamped by low-level correlation between antennas introduced by non-astronomical signals that depend on antenna orientation. Examples include emission from the telescope environment and antenna cross-talk. \citetalias{Keating2015} found that such signals were significant enough to limit the sensitivity if uncorrected. The primary survey data (along with that analyzed in \citetalias{Keating2015}) were observed with a lead-trail strategy to enable removal of ground-correlated emission. To model this contribution, we calculate the variance-weighted average of the three fields within each group as a function of hour angle, for each frequency channel within each baseline. This average is calculate for a singe group of integrations at a time, such that the first 5-second integrations on each field in the group are averaged together to produce a simple model of ground contributions. This model is subtracted from the individual visibilities in each of the three fields. In removing the ground contributions in this fashion, we effectively reduce the number of independent measurements contained within our analysis by 33\%, degrading the sensitivity of our final result by an estimated 18\%.

To evaluate the efficacy of our ground subtraction method, we compare the maximum power measured within our 3D power spectrum (i.e., $\mathcal{P}(\mathbf{k},z)$), before and after the ground subtraction. Prior to ground subtraction, we find that 0.2\% of the individual modes measured exceed the theoretical noise by more than 5$\sigma$, with the largest being $\sim10^{3}$ times greater than expected.  After ground subtraction, we find no modes that exceed $5\sigma$, and the largest of the previously contaminated modes are below $3.5\sigma$.

For the pilot data, observations of different fields were sometimes separated by several days, preventing direct subtraction of the ground contribution. However, a jackknife analysis of the data (discussed further in Section \ref{ssec_jackknife}) indicates that the longer baselines to the outrigger antennas are free of ground contamination and can therefore be included without this subtraction step.

To recover the short-baseline data taken during pilot observations, the data are summed (in the $(u,v,\eta,z)$ domain) across the fields of the pilot survey. A $\chi^{2}$ test is then performed on these data for each position in $u$ and $v$, evaluated across all redshift windows and delay channels (excluding the $\eta=0$ channel). We expect individual data to be thermal-noise dominated; we therefore assume data that exceed the $4\sigma$ confidence threshold for our $\chi^{2}$ test are irreparably contaminated by systematics, and exclude them from further analysis. The remaining data is presumed to be only weakly contaminated by the ground, such that cross-correlation between the pilot data and the ground-subtracted primary data is not expected to be significantly contaminated. With the PSCP method, this cross-correlation only requires using a slightly modified version of Equation \ref{eqn_powspec_pscp}, using the product of the pilot and primary datasets (and dropping the $\mathcal{A}_{k}$ term). For the MLE method, we employ the constraint matrix to downweight the autocorrelation of the pilot data, such that only the autocorrelations of primary survey data and the cross-correlation of pilot and primary survey data contribute to our measurement.
%%%%%% Point Sources%%%%%%
\subsubsection{Point Source Contamination}\label{ssec_psremoval}
\begin{figure}[t]
\begin{center}
\includegraphics[scale=0.5]{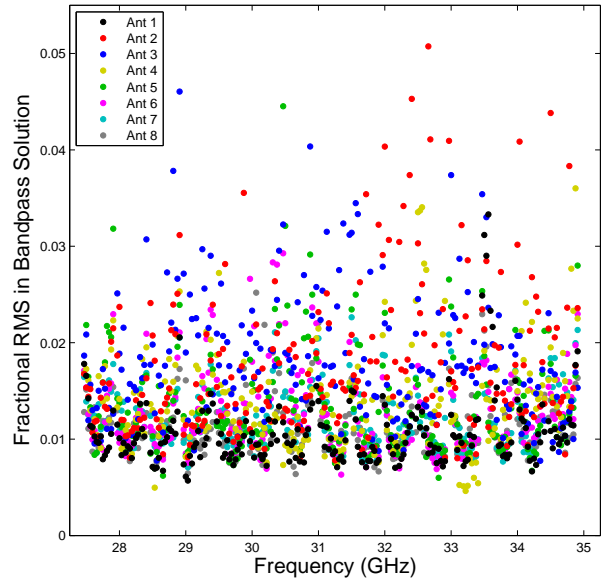}
\caption{The RMS variation in bandpass solutions over the course of the observations. Most channels show 1--2\% variation, with edge channels varying slightly more. 
\label{fig_bpstability}}
\end{center}
\end{figure}
As previously discussed in \citetalias{Keating2015}, the primary source of contamination in our measurement is expected to arise from continuum point source emission. Our primary means of rejecting such contributions is to remove the $\eta=0$ channel, although bandpass calibration errors and non-zero spectral indices of sources will lead to contributions to channels other than $\eta=0$. The power spectrum contribution of continuum point sources can be suppressed further by subtracting the detected sources from the measured visibilities. Our simulations show that removing sources brighter than $\sim0.25$~mJy ($5\times$ the typical image noise) will reduce the contributed power by a factor of 20, corresponding to a residual of $\sim1\ \mu\textrm{K}^2\ h^{-3}\,\textrm{Mpc}^{3}$ 
at $k=1 h\,\textrm{Mpc}^{-1}$, well below the sensitivity achieved in this measurement.

Bandpass calibration errors mix power from the discarded $\eta=0$ channel into the signal channels, even when detected point sources are removed. As shown in Figure \ref{fig_bpstability}, the day-to-day stability of the SZA is excellent -- typically showing an RMS variability of $1.2 \%$ for individual channels. If the errors in bandpass calibration were correlated across days, we expect this level of error to add $100\ \mu\textrm{K}^2\ h^{-3}\,\textrm{Mpc}^{3}$ of power to our measurement. However, our bandpass errors are thermal-noise dominated, and we expect these errors to average down when adding together multiple days worth of data. As bandpass error-related contributions are expected to scale with as the square of the fractional bandpass error, with approximately 400 different bandpass solutions, we expect bandpass errors to only add $1\ \mu\textrm{K}^2\ h^{-3}\,\textrm{Mpc}^{3}$ of power to our present measurement. While this contribution is comparable to the primary contribution from continuum point source for the shortest of baselines, it is well below the sensitivity threshold of our experiment.
%%%%%% Jackknife Tests%%%%%%
\subsubsection{Jackknife Tests}\label{ssec_jackknife}
\begin{figure}[t]
\begin{center}
\includegraphics[scale=0.5]{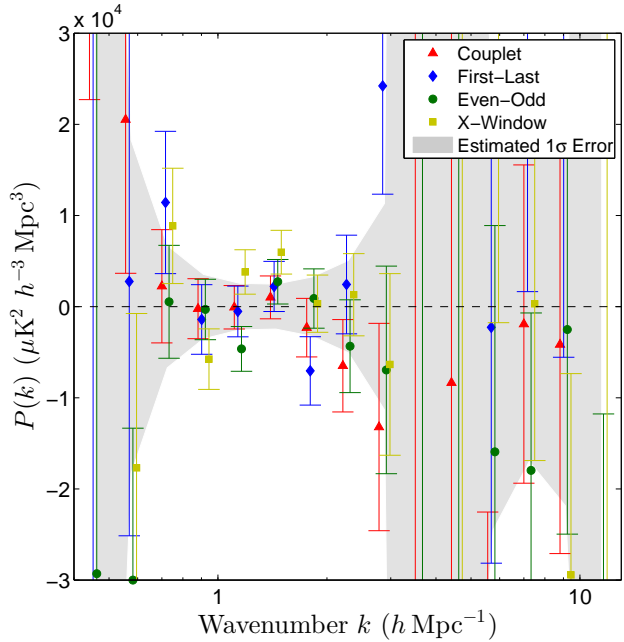}
\caption{Jackknife analysis results for the COPSS data set, along with the estimated noise threshold of our measurement (solid gray). We find the jackknife results are noise-like in  distribution, consistent with the assertion that the data are predominately free of systematics. The largest outlier has $2.4\sigma$ significance, consistent with the expectations of a normally distributed set of data given the $\sim60$ values produced by our jackknife analysis.
\label{fig_jackknife}}
\end{center}
\end{figure}
To verify that our measurement is relatively free of systematic errors that may otherwise contaminate our result, we perform a series of null tests -- referred to as ``jackknife tests'' -- which remove the astronomical signal form the data, via linear combination or randomization of the visibility phases, and search for residual power. There are a total of four jackknife tests performed on the dataset. The  ``couplet'' test differences pairs of time-adjacent visibilities (i.e., for a given frequency channel within the single baseline, the first 5-second integration is subtracted from the second integration). The ``even-odd'' test arranges alternating days' data into two separate sums, then differences these two sums. The ``first-last'' test sums together all data belonging to the first half of the set of days, and differences that with the sum of all data from the second half. The ``cross-window'' test correlates different redshift windows with one another, under the assumption that the CO signal should not correlate between different spectral windows.

The jackknife analysis results (along with their $1\sigma$ errors) are shown in Figure~\ref{fig_jackknife} and in Table~\ref{table_results}. Table~\ref{table_results} also separately presents the results from the  pilot and primary phases of the COPSS experiment. For each sum total result, the probability to exceed (PTE) -- the likelihood of a result of equal or greater statistical significance to be produced by a noise-like event -- is also calculated (based on the values from PCPS method). The results of our jackknife analysis are consistent with noise, suggesting that our final results are not dominated by systematics. We note that for the cross-window test we have correlated windows two steps apart (e.g., window 1 with window 3) due to concerns raised in \citetalias{Keating2015} that adjacent windows may contain a small degree of common noise between them.

We validated our error estimates for the PCSP method through a data randomization test. To do this, we randomized the phases for each day's data before aggregating the data, rendering the signal of interest incoherent over the course of our observations. The power from the randomized data is measured, and the process repeated 100 times to calculate an estimate for the RMS noise power in our measurement. As in \citetalias{Keating2015}, we find that the resultant noise estimates agree with that derived from thermal estimates to within 10\% (i.e., within limits of what we expect given 100 trials). We do find some more significant differences of $\lesssim50\%$ between the two estimates in bins with lowest sensitivity. We attribute this to the smaller number of independent measurements in these bins, which makes their $\chi^2$-distributed amplitudes be less well approximated by a Gaussian distribution.
%%%%%%%%%%%%%%%%%%%%%%%%%%%%%%%%%%%%%%%%%%%%%%%%%%%
%%%%%% Results%%%%%%
\section{Results}\label{sec_results}
\begin{figure*}[t]
\begin{center}
\plottwo{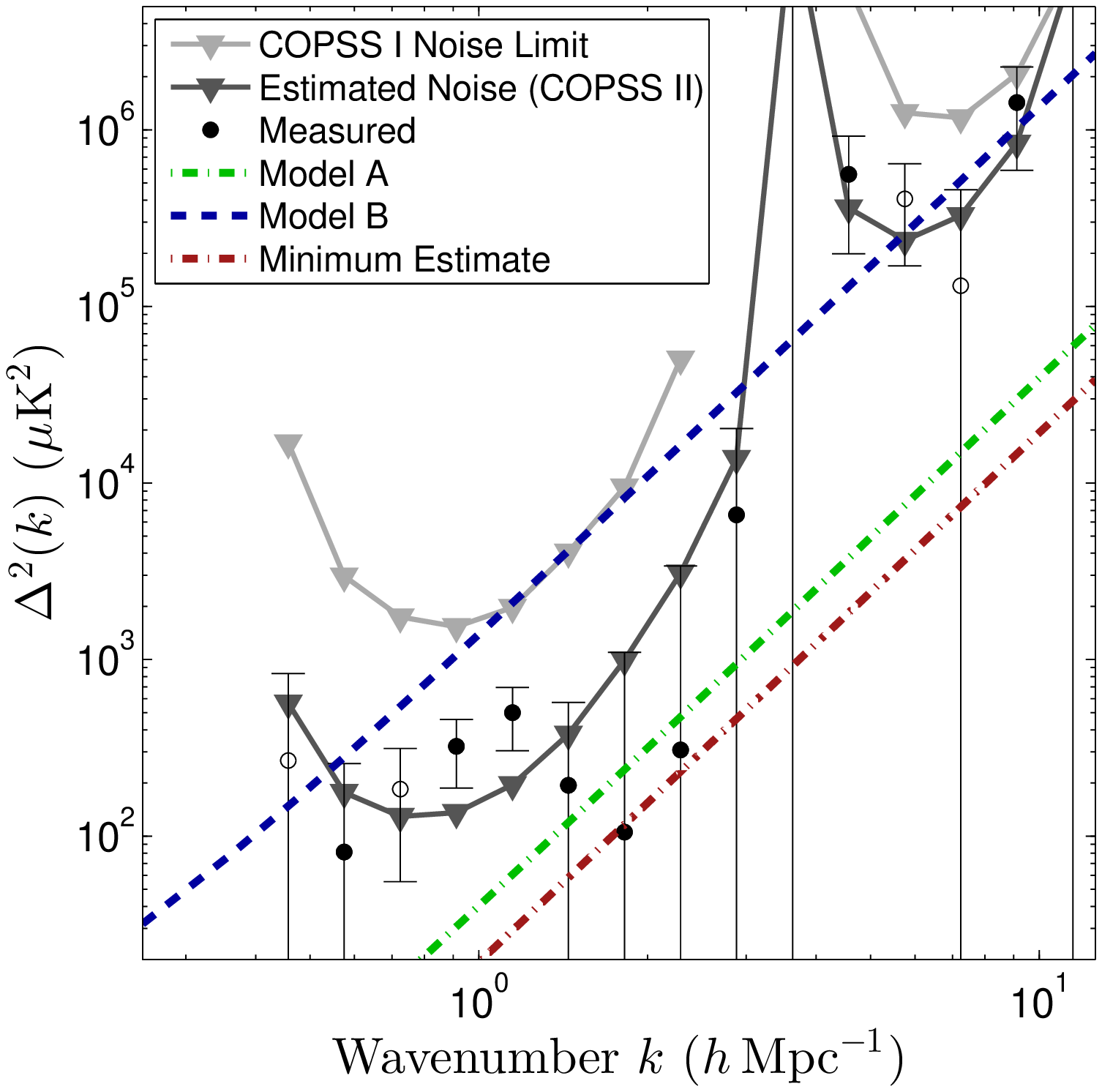}{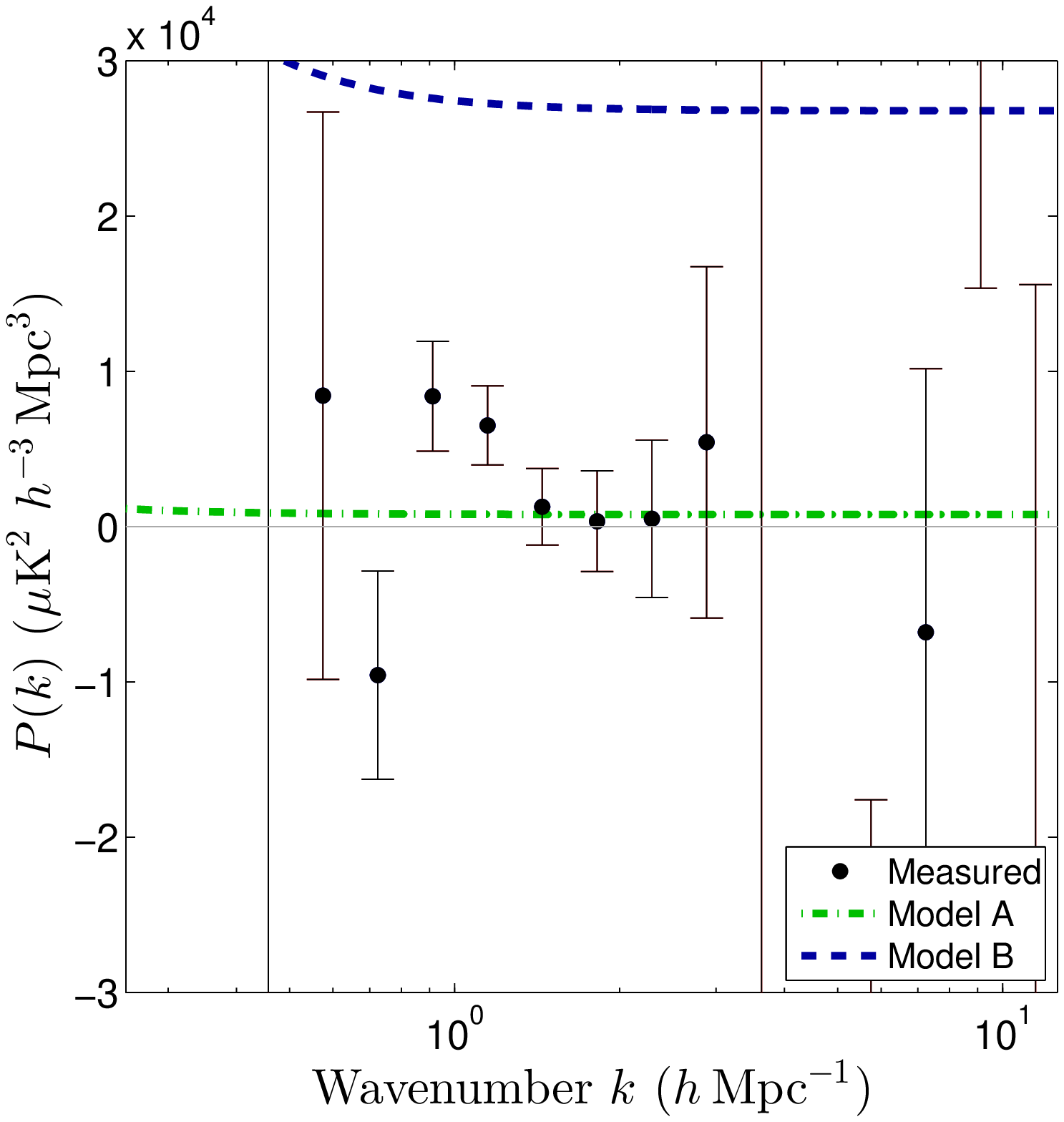}
\caption{\textit{Left}: The result of our power spectrum analysis of the COPSS data set, in the form $\Delta^{2}(k)$. Positive values of $\Delta^{2}(k)$ are shown as filled circles and negative values as open circles, with error bars corresponding to the $1\sigma$ errors on our measured values. Model A (dot-dashed green) and model B (dashed blue) from \cite{Pullen2013} are shown for reference, along with the estimated RMS noise power for this analysis (dark gray triangle) and that of \citetalias{Keating2015} (light gray triangle), absent any astrophysical signal. Also shown is the estimated power that would be contributed by a population of galaxies like those with optical counterparts detected by \citet{Decarli2014}, which provides a lower limit on the power we should observe (in the absence of cosmic variance). \textit{Right}: The power spectrum result, in the form $P(k)$.
\label{fig_finpowspec}}
\end{center}
\end{figure*}

Presented in Table \ref{table_results} and Figure~\ref{fig_finpowspec} are the final results of our analysis of the complete COPSS data set (including those data published in \citetalias{Keating2015}). Our measurement has peak sensitivity at $k = 1.3\ h\,\textrm{Mpc}^{-1}$, with best sensitivity between $k=0.5{-}2\ h \,\textrm{Mpc}^{-1}$ (and marginal sensitivity between between $k=2{-}10\ h \,\textrm{Mpc}^{-1}$). Integrating over all redshift windows and wavenumbers, we detect power of $P_{\textrm{CO}}=3.0_{-1.3}^{+1.3}\times10^{3}\ \mu\textrm{K}^{2} (h^{-1}\,\textrm{Mpc})^{3}$, and reject the null hypothesis ($P_{\textrm{CO}}>0$) at 98.9\% confidence. Placing this measurement into $\Delta^{2}_{N}$ units, where Poisson power grows like $k^3$, requires choosing a $k$ value. At $k = 1\ h\,\textrm{Mpc}^{-1}$, $\Delta^{2}_{\textrm{CO}}=1.5^{+0.7}_{-0.7} \times10^{3}\ \mu\textrm{K}^{2}$.

\begin{figure}[!t]
\begin{center}
\includegraphics[scale=0.5]{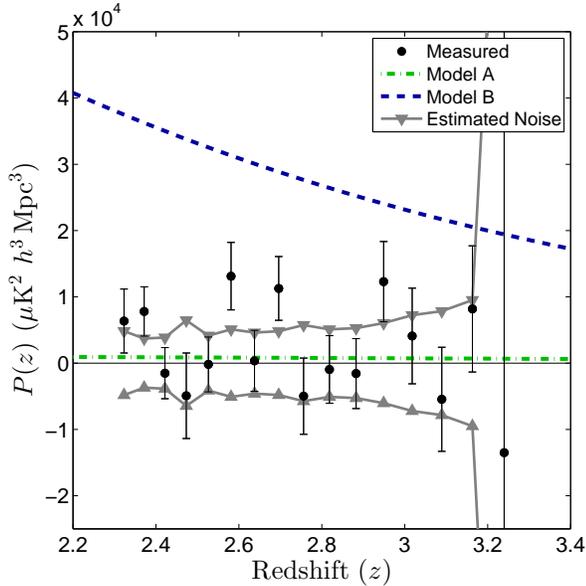}
\caption{The result of our power spectrum analysis as a function of redshift (averaged over all $k$), with corresponding $1\sigma$ errors for each redshift bin. We find that the results for individual bins are consistent with the mean power measured across all bins.
\label{fig_powspecz}}
\end{center}
\end{figure}

Theoretical models (e.g., \citealt{Li2016}, hereafter referred to as \citetalias{Li2016}) suggest that there $P_{\textrm{CO}}$ may evolve significantly over the redshift range sampled by  our measurement ($z=2.3{-}3.3$). We therefore show the results for each individual redshift bin in Figure~\ref{fig_powspecz}. We find weak evidence for decreasing power with increasing redshift: we measure $P_{\textrm{CO}}=4.1_{-1.6}^{+1.6}\times10^{3}\ \mu\textrm{K}^{2} (h^{-1}\,\textrm{Mpc})^{3}$ for the low-redshift half of the data ($z=2.3-2.8$), and $P_{\textrm{CO}}=1.0_{-2.4}^{+2.4}\times10^{3}\ \mu\textrm{K}^{2} (h^{-1}\,\textrm{Mpc})^{3}$ for the high-redshift half of the data ($z=2.8-3.3$). While this trend is not of high enough significance to demonstrate any evolution with redshift, it does agree with the expectation that the measured power should decrease with increasing redshift over the redshift range of our measurement.
\floattable
\begin{deluxetable*}{ccccc|ccc}
\tablecaption{Power Spectrum Measurements.
\label{table_results}}
\tablehead{\colhead{Jackknife Test} & \multicolumn{2}{c}{Primary Data} & \multicolumn{2}{c}{Pilot Data} & \multicolumn{2}{c}{Total} & \colhead{PTE}}
\startdata
           & PCPS         & MLE                 & PCPS        & MLE                 & PCPS        & MLE                 & \\
Couplet    & $-1.1\pm1.4$ & $-1.7^{+1.4}_{-1.4}$       & \phs $2.1\pm3.0$ & \phs$2.4^{+3.0}_{-3.1}$ & $-0.6\pm1.3$ & $-0.9^{+1.3}_{-1.3}$ & 0.62 \\
Even-Odd   & $-2.4\pm1.4$ & $-2.6^{+1.4}_{-1.4}$       & \phs $4.4\pm3.0$ & \phs$4.1^{+3.0}_{-3.0}$ & $-1.1\pm1.3$ & $-1.3^{+1.3}_{-1.3}$ & 0.38 \\
First-Last & \phs$0.6\pm1.6$ & \phs$0.5^{+1.6}_{-1.7}$ & $-1.4\pm3.4$ & $-0.9^{+3.4}_{-3.4}$ & \phs$0.3\pm1.4$ & \phs$0.2^{+1.4}_{-1.4}$ & 0.82 \\
Cross-Win  & \phs$2.5\pm1.5$ & \phs$2.3^{+1.5}_{-1.5}$ & \phs $1.9\pm2.4$ & \phs$2.0^{+2.4}_{-2.4}$ & \phs$2.1\pm1.3$ & \phs$1.9^{+1.3}_{-1.3}$ & 0.11 \\
\hline
Science Result & \phs$3.6\pm1.4$ & \phs $3.3^{+1.5}_{-1.5}$ & $-0.4\pm3.7$ & \phs$0.1^{+3.7}_{-3.8}$ & \phs $3.1\pm1.3$ & \phs $3.0^{+1.3}_{-1.3}$ & 0.01 \\
\enddata
\tablecomments{All power spectrum values are in units of $10^{3}\ \mu\textrm{K}^2\ h^{-3}\,\textrm{Mpc}^{3}$. PTE values are calculated using the PCPS values.}
\end{deluxetable*}

\begin{figure}[!t]
\begin{center}
\includegraphics[scale=0.5]{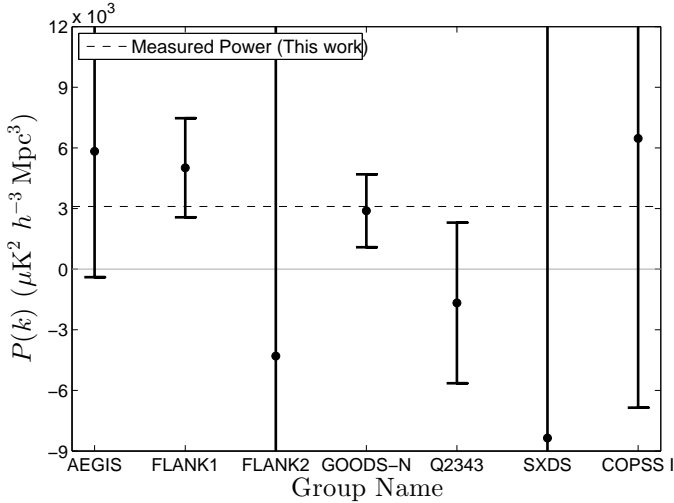}
\caption{Power measured within individual groups of fields. The largest outlier is the Q2343 group, which has power $1.2\sigma$ below what was measured across the entire survey.
\label{fig_trimtests}}
\end{center}
\end{figure}

In Figure~\ref{fig_trimtests}, we consider the measured power in the individual field groups to determine whether any one field dominates our measured power. None of the fields deviates strongly from the average power, with the largest excursion being a 1.2$\sigma$ deficit in the Q2343 group.
%%%%%%%%%%%%%%%%%%%%%%%%%%%%%%%%%%%%%%%%%%%%%%%%%%%
%%%%%% Discussion%%%%%%
\section{Discussion}\label{sec_discussion}
\subsection{Constraints on $A_{\textrm{CO}}$ and $\sigma_{\textrm{CO}}$}\label{ssec_comodels}
\begin{figure*}[t]
\begin{center}
\plottwo{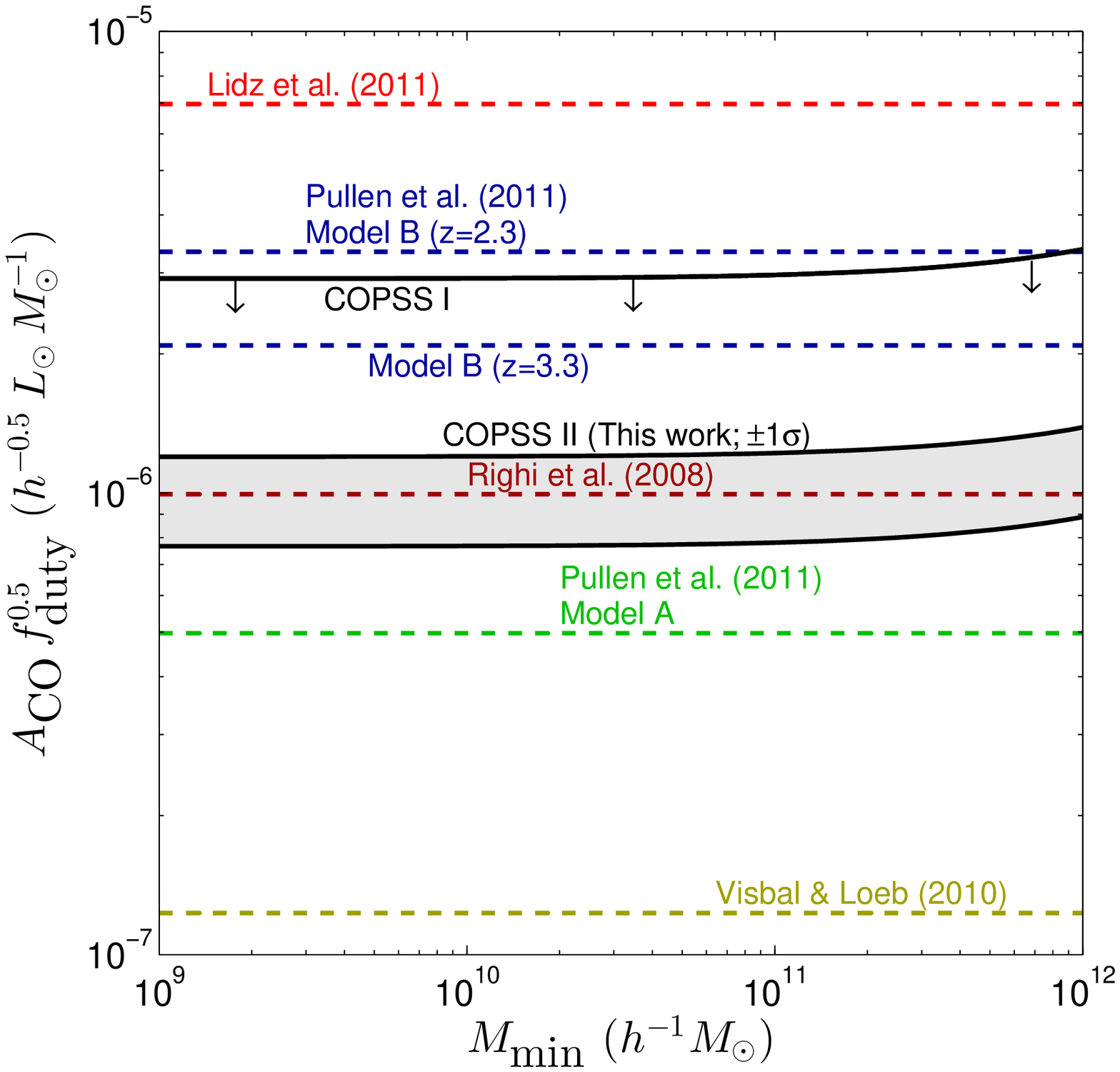}{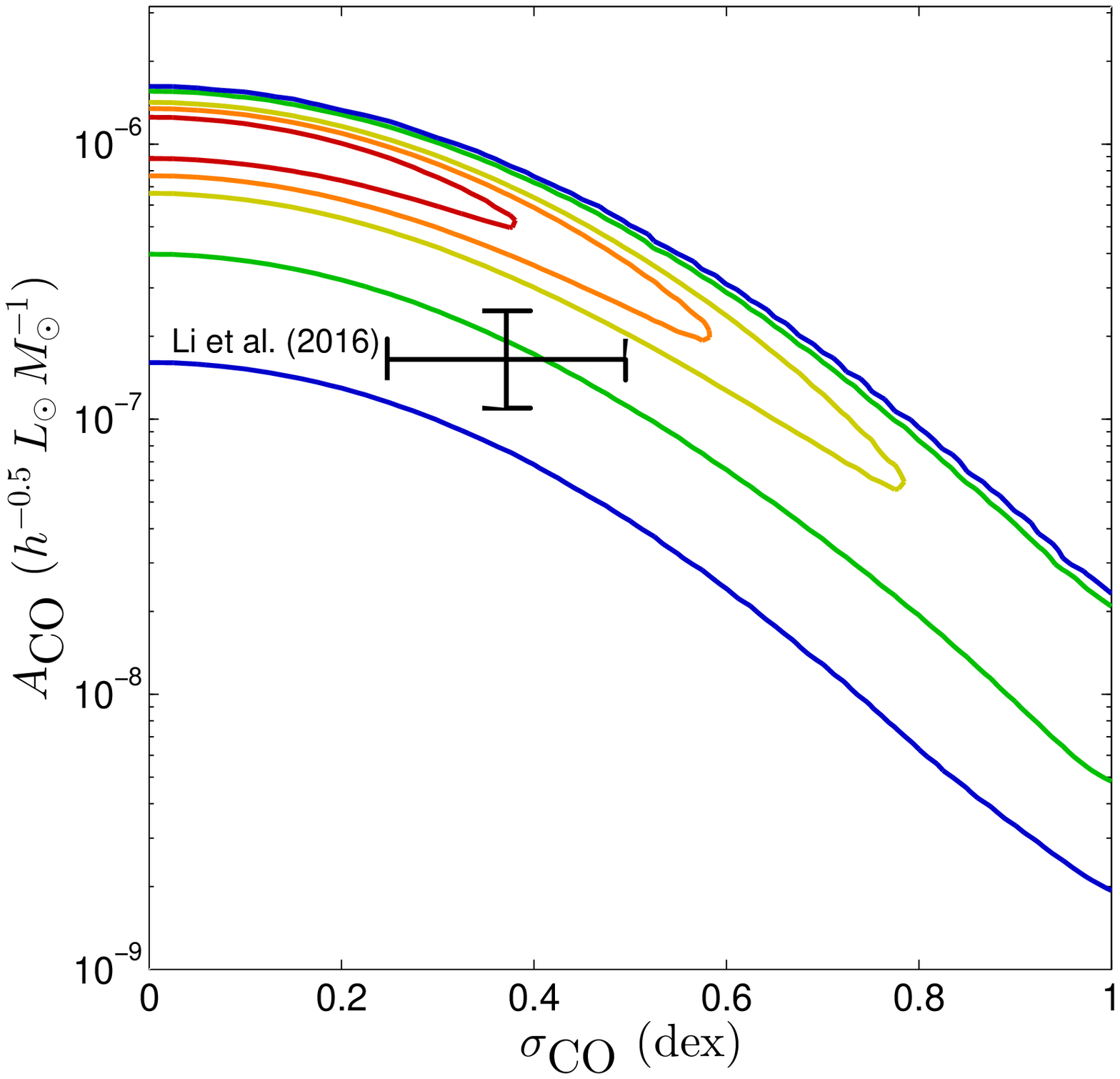}
\caption{\textit{Left}: Constraints on $A_{\textrm{CO}}$ as a function of $M_{\textrm{min}}$. The $1\sigma$ constraints from our analysis (gray) are shown versus theoretical expectations for $A_{\textrm{CO}}$, multiplied by the square root of value for $f_{\textrm{duty}}$ used with each model ($f_{\textrm{duty}}=0.1$ for \citet{Visbal2011}, $f_{\textrm{duty}}=t_{s}/t_{H}$ for all others).\citet{Righi2008} do not explicitly supply a value for $f_{\textrm{duty}}$ or $A_{\textrm{CO}}$ -- we have therefore adopted value of $f_{\textrm{duty}}$ used by the \citet{Pullen2013} and \citet{Lidz2011} models and use the value of $A_{\textrm{CO}}$ determined for this model by \citet{Breysse2014}. \textit{Right}: Constraints on $A_{\textrm{CO}}$ versus $\sigma_{\textrm{CO}}$, with the 25\% (red), 50\% (orange), 68.3\% (yellow), 90\% (green) and 95.4\% (blue) confidence limits shown.  Also shown is the theoretical range of expectations from \citetalias{Li2016} (with $1\sigma$ errors) -- as this model predicts a mass-dependent value for $A_{\textrm{CO}}$, we have used a mean value for $A_{\textrm{CO}}$, weighted by $M^{2}\,dN/dM$ (i.e., the shot power contribution from halos of a given mass).
\label{fig_acoconstraints}}
\end{center}
\end{figure*}
The power spectrum for CO as a function of wavenumber and redshift is given by
\begin{equation}\label{eqn_genpowspec}
P(k,z) = \langle T_{\textrm{CO}} \rangle^{2}b^{2}(z)P_{\textrm{lin}}(k,z) + P_{\textrm{shot}}(z),
\end{equation}
where $T_{\textrm{CO}}$ is the mean brightness temperature, $b(z)$ is the halo bias, $P_{\textrm{lin}}$ is the linear matter power spectrum, and $P_{\textrm{shot}}$ is the shot contribution to the power spectrum. Assuming a linear relationship between CO luminosity ($L_{\textrm{CO}}$) and halo mass, $P_{\textrm{shot}}$ can further be defined as
\begin{multline}\label{eqn_shotnoisemass}
P_{\textrm{shot}}(z) =  \left ( A_{\textrm{CO}} \frac{c^{3}(1+z)^2}{8\pi\nu_{o}^{3}k_{\textrm{B}}H(z)} \frac{L_{\odot}}{M_{\odot}} \right)^{2} \\
f_{\textrm{duty}} \int_{M_{\textrm{min}}}^{\infty} M^{2} \frac{dn(z)}{dM} \,dM,
\end{multline}
where $\nu_{o}$ is the rest frequency of the line, $H(z)$ is the Hubble parameter, $dn(z)/dM$ is the number of halos per unit mass as a function of redshift, $f_{\textrm{duty}}$ is the duty cycle of CO emitters (i.e., the fraction of time a halo hosts CO-emitting galaxies), and $A_{\textrm{CO}}$ is the ratio of CO(1-0) luminosity to halo mass for CO-luminous halos, with units of $L_{\odot}M_{\odot}^{-1}$  \citep{Lidz2011,Breysse2014}. Halos with masses below the low-mass limit, $M_{\textrm{min}}$, are assumed to lack sufficient CO, and thus do not appreciably contribute to the larger-scale emission detectable in our power spectrum measurement (e.g., \citealt{Visbal2010,Lidz2011}). Figure~\ref{fig_acoconstraints} shows  that our result is not sensitive to the choice of this parameter.

As discussed in \citetalias{Keating2015}, several models adopt $f_{\textrm{duty}}=t_{*}/t_\textrm{age}$, where $t_{\textrm{age}}$ is the Hubble time and $t_{*}\approx100$ Myr is the timescale of star formation. Under these assumptions, this sets $f_{\textrm{duty}}\approx0.05$ for $z\sim3$ -- much lower than the near-unity values typically observed \citep{Noeske2007,Lee2009,Tacconi2013}. This tension makes the $f_{\textrm{duty}}$ parameter problematic over the redshift range considered in our analysis. We instead introduce a term analogous to that found in \citetalias{Li2016}: $\sigma_{\textrm{CO}}$, the log-scatter (in units of dex) of the correlation between halo mass and $L_{\textrm{CO}}$. This parameter allows for the realistic possibility that CO luminosity does not map directly to halo mass, but instead may scatter significantly, and assumes a Gaussian form to this scatter in logarithmic units. A large value of $\sigma_\textrm{CO}$ implies a weak correlation between luminosity and halo mass.  We note that \citetalias{Li2016} disaggregate this term into the scatter in the underlying relationships in the halo mass to CO luminosity correlation (labeling the aggregate term as $\sigma_{\textrm{tot}}$). As we are unable to separate those different sources of scatter in our measurement, we have chosen to use a single, aggregate term in our analysis. Accounting for this scatter requires a minor modification to Equation \ref{eqn_shotnoisemass}, replacing the $f_{\textrm{duty}}$ term with $p_{\sigma}$, defined as the fractional change in shot power induced by $\sigma_{\textrm{CO}}$. We further define $p_{\sigma}$ as 
\begin{equation}\label{eqn_shotnoise_sigma}
p_{\sigma} = \int_{-\infty}^{\infty}{\frac{10^{2x}}{\sqrt{2\pi\sigma^{2}_{\textrm{CO}}}}e^{(-x^{2}/2\sigma_{\textrm{CO}}^{2})}} \, dx.
\end{equation}
We note that as we have defined it in Equation \ref{eqn_shotnoise_sigma}, $p_{\sigma}$ will always be greater than or equal to unity (with $p_{\sigma}$ monotonically increasing with $\sigma_{\textrm{CO}}$), and non-zero values of $\sigma_{\textrm{CO}}$ therefore imply higher $P_{\textrm{CO}}$ than otherwise expected for a given $A_{\textrm{CO}}$. This is in contrast with $f_{\textrm{duty}}$, which should always be less than or equal to unity (and therefore implies a \emph{lower} $P_{\textrm{CO}}$ than otherwise expected for a given $A_{\textrm{CO}}$).

We show our constraints on $A_{\textrm{CO}}$ in Figure \ref{fig_acoconstraints}, as a function of $M_{\textrm{min}}$ (for those models dependent upon $f_{\textrm{duty}}$), and as a function of $\sigma_{\textrm{CO}}$. As our constraint on $A_{\textrm{CO}}$ is weakly dependent on $M_{\textrm{min}}$, for those constraints which are a function of $\sigma_{\textrm{CO}}$, we adopt a value of $M_{\textrm{min}}=10^{10}\ M_{\odot}$ (noting that any choice of $M_{\textrm{min}}<10^{11}\ M_{\odot}$ has a near-negligible effect on our calculations). Under their model, \citetalias{Li2016} adopt a value of $\sigma_{\textrm{CO}}\approx0.37\pm0.12$ dex -- over the 95\% confidence interval of $\sigma_{\textrm{CO}}$, the constraint on $A_{\textrm{CO}}$ changes by as much as a factor of 40. This is the manifestation of the degeneracy between $A_{\textrm{CO}}$ and $p_{\sigma}$ in the combination of Equations \ref{eqn_shotnoisemass} and \ref{eqn_shotnoise_sigma} (although we weakly break this degeneracy by measuring the power of multiple groups of widely separated fields). Additional data at smaller $k$, where the clustering component of the power spectrum dominates, would help break this degeneracy because the clustering component is less affected by $\sigma_{\textrm{CO}}$. In the absence of constraints on $\sigma_{\textrm{CO}}$, we are restricted to placing only an upper limit on the halo mass to CO luminosity ratio, with $A_{\textrm{CO}}<1.5\times10^{-6}\ L_{\odot}\ M_{\odot}^{-1}$ (corresponding to the 68\% confidence limit where $\sigma_{\textrm{CO}}=0$). However, if we adopt the value of $\sigma_{\textrm{CO}}$ from \citetalias{Li2016} (marginalizing over the uncertainty in this parameter), we determine $A_{\textrm{CO}}=6.3_{-2.5}^{+4.0}\times10^{-7}\ L_{\odot}\ M_{\odot}^{-1}$.

One can also use a variant of Equation \ref{eqn_shotnoisemass} to convert ``blind'' detections of CO emitters into an estimate for the minimum power. For this estimate, we will utilize those detections presented in \citet{Decarli2014}. We consider only those galaxies with optical counterparts, as emitters without counterparts are more prone to being either spurious detections or incorrectly ascribed to the wrong redshift (by incorrectly identifying which rotational transition is being observed). Assuming $L^{\prime}_\textrm{{CO(3-2)}}/L^{\prime}_\textrm{{CO(1-0)}}=0.5$ \citep{Walter2014}, we estimate the minimum shot power to be $P_{\textrm{CO,min}}=3.8_{-1.1}^{+3.7}\times10^{2}\ \mu\textrm{K}^{2} (h^{-1}\,\textrm{Mpc})^{3}$. We note that this minimum estimate resides below the power detected in our analysis, and resides near the bottom edge of the 95\% confidence range for $P_{\textrm{CO}}$ (as presented in Section \ref{sec_results}). 

Our present constraints lie below predictions of Model B from \citet{Pullen2013} (as well as \cite{Lidz2011}, which was previously excluded in \citetalias{Keating2015}). These constraints also lie well above predictions made by \citet{Visbal2010}, as does our estimate for the minimum shot power. We now briefly consider what we can learn from the exclusion of these models. At $z\sim3$, the \citeauthor{Visbal2010} model likely suffers from the fact that it is tailored for $z\geq6$, and predicts global star formation rates (SFRs) that are a factor of a few different than what is observed at $z\sim3$, although this difference alone does not completely account for the discrepancy between this model and our lower limit. \citeauthor{Visbal2010} also use M82 as a template to calibrate their SFR-$L_\textrm{CO}$ relationship \citep{Weiss2005}, which predicts CO luminosities that are a factor of a few below what is observed in massive main-sequence galaxies at $z\sim2$ \citep{Tacconi2013}. Our present constraints on $P_{\textrm{CO}}$ suggest that molecular gas properties of local starburst galaxies (like M82) are not well-matched to those of normal galaxies at $z\sim3$ (though they may still hold for higher redshift galaxies).

Model B of \citet{Pullen2013} uses the SFR function parameters from \citet{Smit2012} and a prescription for the SFR-$L_{\textrm{CO}}$ relationship observed in local galaxies \citep{Kennicutt1998,Wang2011} to calculate an estimate for the mean brightness temperature of CO, and then uses this to scale $A_{\textrm{CO}}$ from the value derived in their model A. However, this adjustment overpredicts the number of extremely CO-luminous objects (i.e., $L_{\textrm{CO}}\geq10^{8}L_{\odot}$). As both models A and B are calibrated against the global SFR and mean CO brightness temperature (respectively), the inclusion of $f_{\textrm{duty}}$ within these models requires that a small subset (i.e., $f_{\textrm{duty}}$) of all halos have enhanced emission ($f_{\textrm{duty}}^{-1}$) in order to be consistent with the global values. The lack of detection of such objects suggests that either the Model B estimate for $A_{\textrm{CO}}$ is too high, or that $f_{\textrm{duty}}$ is near-unity (as other observational evidence suggests).
\subsection{Constraints on the CO Luminosity Function}\label{ssec_columfunc}
\begin{figure*}[t]
\begin{center}
\includegraphics[scale=0.425]{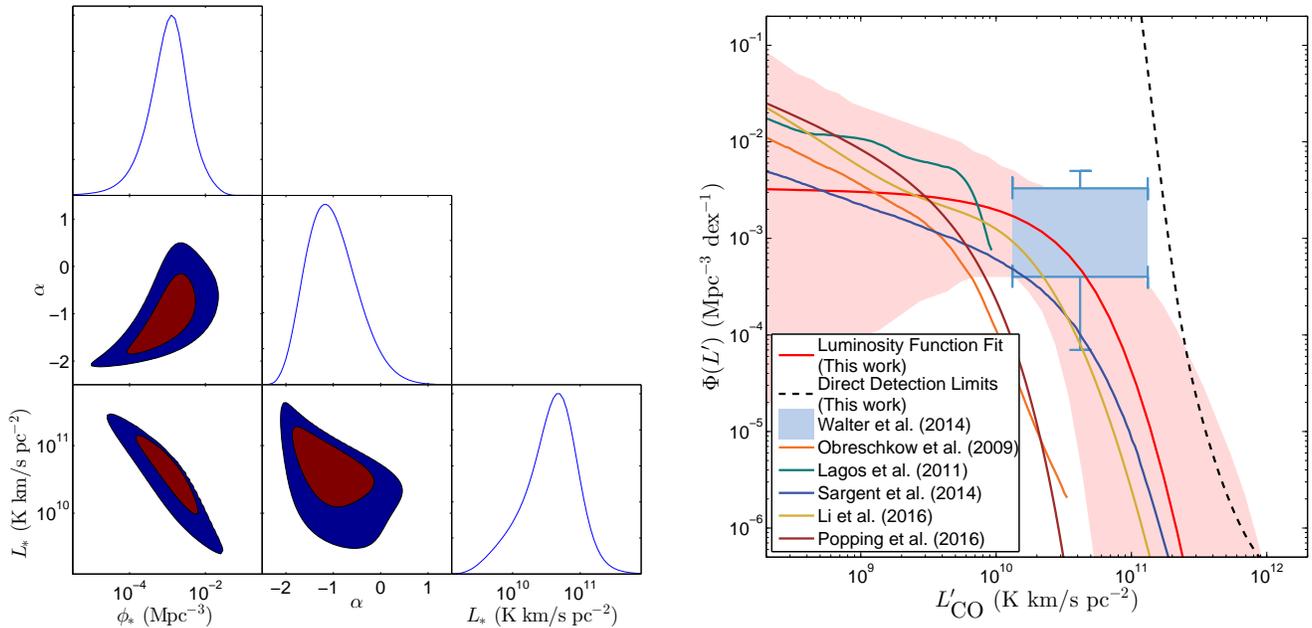}
\caption{\textit{Left}: The constraints on the individual Schechter parameters of the luminosity function: $\phi_{*}$, $\alpha$ and $L_{*}$.  \textit{Right}: The fitted luminosity function and $1\sigma$ errors (red), along with the direct detection constraints from the COPSS data set (dashed), and the constraint on the CO luminosity function constraint at $z=2.75$ (light blue) from \cite{Walter2014}. Shown for comparison are models from \citet{Obreschkow2009e} (orange), \citet{Lagos2011} (dark green), \citet{Sargent2014} (dark blue), \citet{Li2016} (yellow) and \citet{Popping2016} (brown).
\label{fig_columfunc}}
\end{center}
\end{figure*}
Theoretical models indicate that our measurement should sample the shot noise portion of the CO power spectrum. Accordingly, we measure the second moment of the CO luminosity function at $z\sim3$. The second moment of the luminosity function, $\int L^{2} \Phi(L)\,dL$, is related to the shot power by 
\begin{equation}\label{eqn_powtolsq}
P_{\textrm{shot}}(z) = \left( \frac{c^{3}(1+z)^2}{8\pi\nu_{o}^{3}k_{\textrm{B}}H(z)} \right)^{2} \int L^{2} \Phi(L)\,dL.
\end{equation}

One can use the value for the second moment, in combination with data from direct detection efforts, to place constraints on the shape of the luminosity function. To do so, we will assume that the luminosity function is (to first order)  well-described by the Schechter function \citep{Schechter1976}, which has the general form
\begin{equation}\label{eqn_schechter}
\Phi(L) dL = \phi_{*} \left (\frac{L}{L_{*}} \right )^{\alpha} e^{-L/L_{*}} dL/L_{*}.
\end{equation}
Equation \ref{eqn_schechter} is nominally parameterized by a high-luminosity cutoff, $L_{*}$, a low-luminosity power law index, $\alpha$, and a normalization factor for the overall density of luminous sources $\phi_{*}$. For our analysis, we evaluate the likelihood of the combined choice of these three parameters parameters by evaluating the second moment of the luminosity function produced. We will further weight this likelihood by 
\begin{enumerate}
\item the galaxies detected in CO(3-2) with optical counterparts by \citet{Decarli2014},
\item a lack of detections of individual emitters within the COPSS dataset of $\geq5\sigma$ significance within twice the FWHM of the primary beam, and
\item a prior on the slope of low-luminosity end of the luminosity function.
\end{enumerate}
The search for individual emitters within our data was performed assuming a Gaussian emission profile (of width $\Delta v = 300\ \textrm{km}\ \textrm{s}^{-1}$, consistent with observations of \citealt{Decarli2014}). Due to the relatively coarse channelization, the search for individual emitters was conducted by searching only single and 2-channel averaged maps for any points above a threshold of $5\sigma$. Under the Schechter parameterization, our measurement is generally more sensitive to changes in $\phi_{*}$ and $L_{*}$, and less sensitive to changes in $\alpha$. We provide a loose prior of $\alpha=-1.5\pm0.75$ for this parameter based on the SFR function parameters derived at $z\sim4$ in \citet{Smit2012}, 
based on the observed linear relationship between SFR and CO luminosity at high redshift \citep{Tacconi2013}.

In including data from \citet{Decarli2014}, we consider only those galaxies with optical counterparts, as emitters without counterparts are more prone to being either spurious detections or incorrectly ascribed to the wrong redshift (by incorrectly identifying which rotational transition is being observed). In evaluating the likelihood of any set of parameters for the luminosity function, we weight each particular parameter by $\mathcal{L}_{\textrm{gal}}$, the likelihood of observing \emph{at least} the number of objects detected in any particular survey. We further define $\mathcal{L}_{\textrm{gal}}$ as
\begin{equation}\label{eqn_ngallikelihood}
\mathcal{L}_{\textrm{gal}}=1-\sum \limits_{n=n_{\textrm{gal}}+1}^{\infty} Pois(n;\textrm{V}_{z}\rho_{\textrm{gal}})
\end{equation}
In Equation \ref{eqn_ngallikelihood}, $n_{\textrm{gal}}$ is the number of galaxies detected within a particular bin, $\rho_{\textrm{gal}}$ is the expected number density of galaxies (based on the set Schechter parameters being evaluated), and $Pois(k;\lambda)$ is the probably of detecting $k$ objects given a Poisson distribution with mean $\lambda$.

The results of our likelihood analysis are shown in Figure \ref{fig_columfunc}. With our data (along with the constraints and priors mentioned earlier), we constrain $\phi_{*}=1.3^{+0.6}_{-0.7}\times10^{-3}\ L_{\odot}^{-1}\ \textrm{Mpc}^{-3}$ and $L_{*}=4.5_{-1.9}^{+1.4}\times10^{10}\ \textrm{K}\ \textrm{km}\ \textrm{s}^{-1}\ \textrm{pc}^{-2}$ to 68\% confidence. We use these constraints, along with importance sampling, to generate a fit (and $1\sigma$ errors) on the CO luminosity function at $z\sim3$. Our fitted luminosity function agrees with earlier constraints made by \citet{Walter2014}, as well as model predictions made in \citetalias{Li2016} and \citet{Sargent2014}. Our fitted function also appears to disfavor those model predictions made by \citet{Obreschkow2009b}, \citet{Lagos2011}, and \citet{Popping2016}, in that all three appear to underpredict the number of higher luminosity objects at $z\sim3$.
\subsection{Constraints on Cosmic Molecular Gas Abundance}\label{ssec_molgas}
\begin{figure}[t]
\begin{center}
\includegraphics[scale=0.5]{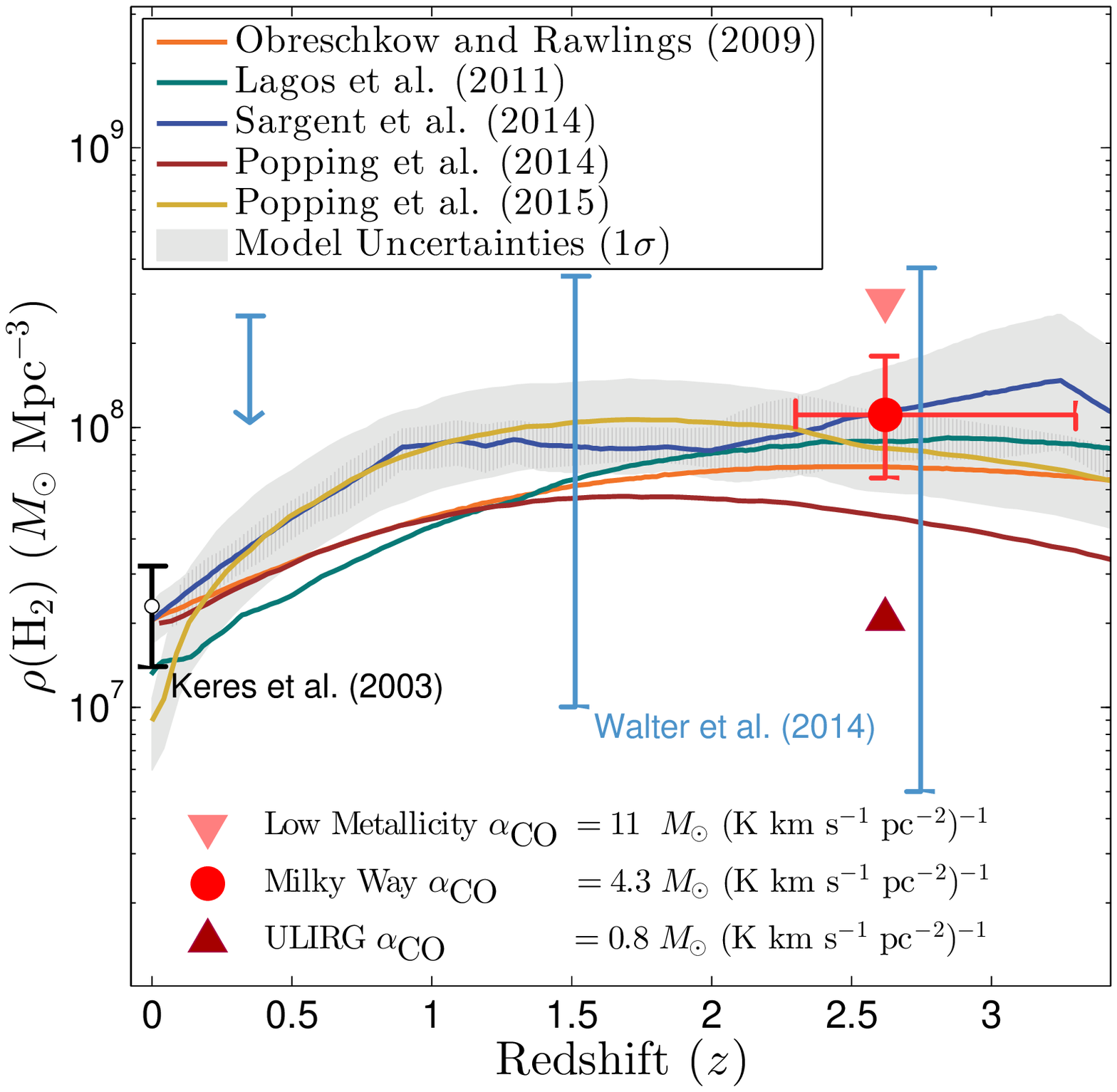}
\caption{The constraint on the cosmic molecular gas density. Shown are the estimates and 68\% confidence region for $\rho{\textrm{H}_{2}}$ adopting a Milky Way-like value for $\alpha_{\textrm{CO}}$ (red circle), as well as the estimate assuming an $\alpha_{\textrm{CO}}$ appropriate for ULIRGs (dark red upward-pointing triangle) and for low-metallicity main-sequence galaxies (pink downward-pointing triangle). Shown for comparison are the constraints from \citet{Walter2014} (light blue). Also shown are the theoretical expectations from \citet{Obreschkow2009c} (orange) and \citet{Lagos2011} (green), \citet{Sargent2014} (dark blue), \citet{Popping2014} (brown; uses modeling similar to that of \citealt{Popping2016}), and \citet{Popping2015} (yellow; which uses modeling of \citet{Behroozi2013} match halo mass to various galaxy properties, similar to what was used for \citetalias{Li2016}). Shown in gray are the 68\% confidence limits on the \citeauthor{Sargent2014} and \citet{Popping2015} models.
\label{fig_rhoh2}}
\end{center}
\end{figure}
Assuming a linear relationship between $L_{\textrm{CO}}$ and $M_{\textrm{H}_{2}}$, one can use the CO luminosity to molecular gas mass conversion factor, $\alpha_{\textrm{CO}}$, to estimate the mass fraction (with respect to the halo mass) of the molecular gas within galaxies, $f_{\textrm{H}_{2}}$. For a Milky Way-like $\alpha_{\textrm{CO,MW}}=4.3 \ M_{\odot}\ (\textrm{K}\ \textrm{km}\ \textrm{s}^{-1}\ \textrm{pc}^{-2})^{-1}$ \citep{Frerking1982,Dame2001} -- equivalent to $8.7\times10^{4}\ M_{\odot}\ L_{\odot}^{-1}$ \citep{Solomon1992} -- our constraint on $A_{\textrm{CO}}$ translates to a limit on the molecular gas mass fraction of $f_{\textrm{H}_{2}}=5.5^{+3.4}_{-2.2}\times10^{-2}$. This constraint applies to halos with mass $\sim10^{12}M_{\odot}$, which dominate the power spectrum for the scales we measure.

To translate this $f_{\textrm{H}_{2}}$ into a constraint on the cosmic $\textrm{H}_{2}$ density ($\rho(\textrm{H}_{2})$), we note that many works have found that $f_{\textrm{H}_{2}}$ peaks around halo masses of $\sim10^{12}\ M_{\odot}$ (e.g., \citealt{Popping2015,Lagos2011}). For our estimate we assume a linear decrease in $f_{\textrm{H}_{2}}$ with $M/M_{0}$ below $M_{0}=5\times10^{11}$, as indicated by \citet{Popping2015}. This eliminates the need to arbitrarily choose a minimum halo mass that contains molecular gas. While \citet{Popping2015} also see a linear decease in $f_{\textrm{H}_{2}}$ above $10^{12}\ M_{\odot}$, we find that including such a variation has minimal effect on our conclusions. Using this prescription, adopting the fiducial value from \citetalias{Li2016} of $\sigma_{\textrm{CO}}=0.37\pm0.12$, and integrating over the halo mass function of \citet{Tinker2008}, we find $\rho_{z\sim3}(\textrm{H}_{2})= 1.1^{+0.7}_{-0.4} \times10^{8}\ M_{\odot}\ \textrm{Mpc}^{-3}$, as shown in Figure~\ref{fig_rhoh2}. Several theoretical predictions for the cosmic molecular gas density at $z\sim3$ lie within the 68\% confidence interval of our constraint, although the theoretical predictions of \citet{Popping2014} do lie outside of this interval. Combined with the disagreement between our fitted CO(1-0) luminosity function and the model of \citet{Popping2016} (which shares a very similar framework to that of \citealt{Popping2014}), this suggests that this particular model may be underestimating the molecular gas abundance within galaxies at $z\sim3$ (further discussed in \citealt{Popping2015}).

An alternate method for calculating $\rho(\textrm{H}_{2})$ is to find the volume emissivity of the CO(1-0) transition -- using our fit for the luminosity function found in Section~\ref{ssec_columfunc} -- and applying an appropriate choice of $\alpha_{\textrm{CO}}$ to this value (similar to what was done in \citetalias{Keating2015}). This method does have some limitations, particularly that it is very sensitive to the faint-end slope ($\alpha$) in the luminosity function, which we constrain primarily with external data. Nevertheless, we find that integrating the CO(1-0) luminosity function for $L^{\prime} \geq 10^{8}\ \textrm{K}\ \textrm{km}\ \textrm{s}^{-1}\ \textrm{pc}^{-2}$ (which corresponds to the $M_{\textrm{min}}$ used to determine $A_{\textrm{CO}}$ in Section~\ref{ssec_comodels}) and adopting the Milky Way value of $\alpha_{\textrm{CO}}$, we find $\rho_{z\sim3}(\textrm{H}_{2})= 1.2 \times10^{8}\ M_{\odot}\ \textrm{Mpc}^{-3}$ . This agrees with our primary estimate to better than 10\% and is well within within the range of the estimated errors.
\subsubsection{Systematic Uncertainty in the Molecular Gas Abundance}\label{ssec_alphaco}
Our estimate of $\rho(\textrm{H}_{2})$ is linearly dependent on our choice of $\alpha_{\textrm{CO}}$. We have adopted $\alpha_{\textrm{CO,MW}}$ based in part on the model of \citet{Sargent2014}, which suggests that this value is appropriate for the galaxies that we expect to dominate our power spectrum measurement (i.e., those hosted by $10^{12}\ M_{\odot}$ halos). Other models predict much larger values of $\alpha_{\textrm{CO}}$ for our high-redshift population (e.g., \citealt{Genzel2015}). We therefore briefly consider the range of values for $\alpha_{\textrm{CO}}$ indicated by past work, and the corresponding effect on our estimate of $\rho(\textrm{H}_{2})$.

With their limited star formation histories, high redshift galaxies ($z\gtrsim2$) typically possess gas-phase metallicities lower than that of the Milky Way, leading to molecular gas that is relatively CO-poor \citep{Bolatto2013,Carilli2013}. To estimate the what impact the metallicity of galaxies may have on $\alpha_{\textrm{CO}}$ (and by extension, $\rho(\textrm{H}_{2})$), we again consider that emission from galaxies with halo masses of $10^{12}$ are expected to dominate our measurement. \citet{Behroozi2013} found that halos of this mass should be associated with stellar masses of $M_{\star}\sim10^{10.5}\ M_{\odot}$ and star formation rates of $SFR\sim30\ M_{\odot}\ \textrm{yr}^{-1}$ at $z\sim3$. Measurements of the relationship between stellar mass and metallicity indicate that such galaxies should have gas-phase metallicities of $Z\sim0.5\ Z_{\odot}$ \citep{Mannucci2010,Troncoso2014,Onodera2016,Steidel2016}. We can combine these parameters with the empirical formula for $\alpha_{\textrm{CO}}$ from \citet{Genzel2015} to estimate $\alpha_{\textrm{CO}}\sim11\ M_{\odot}\ (\textrm{K}\ \textrm{km}\ \textrm{s}^{-1}\ \textrm{pc}^{-2})^{-1}$. This value would increase our estimate of $\rho_{z\sim3}(\textrm{H}_{2})$ by nearly a factor of 3, which would place it substantially above the theoretical models shown in Figure \ref{fig_rhoh2}. If the \citet{Genzel2015} $\alpha_{\textrm{CO}}$ model is correct, our data suggest that models have significantly underestimated the molecular gas density at $z\sim3$. However, the high-redshift calibration of this relation is based on a small number of relatively massive, optically selected galaxies, which may not represent the population captured by our measurement.

At the other extreme for $\alpha_{\textrm{CO}}$, we consider a scenario where our measurement is dominated by dusty star-forming galaxies (DSFGs) with high star formation rates. These galaxies have been found to have much lower values of $\alpha_{\textrm{CO}}\sim1-2 \ M_{\odot}\ (\textrm{K}\ \textrm{km}\ \textrm{s}^{-1}\ \textrm{pc}^{-2})^{-1}$ (e.g., \citealt{Sargent2014,Spilker2015}), comparable to the value of 0.8 found for nearby ultra-luminous infrared galaxies (ULIRGs; \citealt{Downes1998}). Adopting this value would drop $\rho_{z\sim3}(\textrm{H}_{2})$ by a factor of more than 5, to a point well below the models in Figure~\ref{fig_rhoh2}. However, such a low value of $\alpha_{\textrm{CO}}$ is at odds with the trends toward higher $\alpha_{\textrm{CO}}$ with reduced metallicity expected from first principles and observed at low and moderate redshift \citep{Wolfire2010,Genzel2015}.

We also note that in addition to $\alpha_{\textrm{CO}}$, our choices of $M_{0}$, $\sigma_{\textrm{CO}}$, and the precise scaling relationship between halo mass and molecular gas mass fraction also impact our $\rho(\textrm{H}_{2})$ estimate. Our analysis suggests that of these factors, the uncertainty in value of $\sigma_{\textrm{CO}}$ has the strongest impact on our estimate -- over the 95\% confidence interval of our prior for $\sigma_{\textrm{CO}}$, our estimates for $\rho_{z\sim3}(\textrm{H}_{2})$ differ by as much as a factor of $\sim4$.
\subsection{Cosmic Variance and Limits of Significance}\label{ssec_cosmicvar}
We now consider the impact of cosmic variance on our measurement. In the shot-power regime, there are two sources of cosmic variance: large-scale structure inducing a local over/underdensity over the area of our measurement \citep{Tegmark1998}, and the Poisson noise associated with the limited number of massive halos (which are the primary contributors to the shot power). The power measured is roughly proportional to the number density of emitters within the volume measured -- specifically, the number density of ``luminous-but-common'' emitters that are contributing most to the shot component of the power spectrum. With a total survey volume of $4.9\times10^{6}\ h^{-3}\, \textrm{Mpc}^{3}$, this amounts to $\sim10{^4}$ halos with masses of order $10^{12}\ M_{\odot}$, translating to cosmic variance-induced errors of a few percent in our measurement (which is a factor of several greater than that induced by large-scale structure over our survey area). However, if one assumes that $\sigma_{\textrm{CO}}\ne0$, then the scatter of the halo mass to CO luminosity relationship will have the effect of reducing the number of halos that appreciably contribute to the power measurement (i.e., a subset of this population will become slightly more luminous, and hence will contribute more to the shot power). Adopting a fiducial value of $\sigma_{\textrm{CO}}=0.37$ and assuming a linear scaling between halo mass and $L_{\textrm{CO}}$, we estimate cosmic variance-induced errors of $\Delta P/P\approx 0.17$ in our measurement. As this contribution is insignificant in comparison to thermal noise estimates, we find that the impact of cosmic variance is minor. We have therefore neglected cosmic variance in our power spectrum estimates, though have included its impact in our calculations for $A_{\textrm{CO}}$, $\rho(\textrm{H}_{2})$, and the CO luminosity function parameters.
%%%%%%%%%%%%%%%%%%%%%%%%%%%%%%%%%%%%%%%%%%%%%%%%%%%
%%%%%% Conclusion%%%%%%
\section{Conclusion}\label{sec_conclusion}
In this paper, we have constrained the power spectrum for CO at $z\sim3$ to $P_{\textrm{CO}}=3.0_{-1.3}^{+1.3}\times10^{3}\ \mu\textrm{K}^{2} (h^{-1}\,\textrm{Mpc})^{3}$, or $\Delta^{2}_{\textrm{CO}}(k\! = \! 1 \ h\,\textrm{Mpc}^{-1})=1.5^{+0.7}_{-0.7} \times10^{3}\ \mu\textrm{K}^{2}$. We have used this to constrain the relationship between halo mass and CO luminosity, to place limits on the CO luminosity function, and to estimate the cosmic molecular gas density at $z\sim3$. We exclude Model B from \citet{Pullen2013}, as well as the model from \citet{Visbal2010}.

Upcoming 3mm observations with the Yuan-Tseh Lee Array \citep{Bower2015,Ho2009} will offer increased sensitivity and will be capable of deeply probing the CO power spectrum at $z\sim3$. As the Lee Array and the SZA share similar spatial frequency and redshift coverage, the combination of these observations will also enable an opportunity for cross-correlation between the CO(1-0) and CO(3-2) transitions, offering both improved sensitivity and a more complete probe into the physical properties of the molecular gas fueling early star formation. These observations will also be sensitive to lower wavenumbers, where contributions from the clustering of galaxies are more likely to dominate the power spectrum. The added constraints on the cluster-power contributions to the power spectrum will be vital in constraining $A_{\textrm{CO}}$ and $\sigma_{\textrm{CO}}$, and will offer added insight into the population sub-$L_{*}$ CO emitters.
%%%%%%%%%%%%%%%%%%%%%%%%%%%%%%%%%%%%%%%%%%%%%%%%%%%
%%% Acknowledgments%%%%%%
\acknowledgments
The authors would like to thank the referee for their thoughtful and timely feedback, which helped improve the quality and clarity of this manuscript. We thank D. Hawkins, J. Lamb, D. Woody, and S. Muchovej for their technical support. We would also like to thank C. Heiles, R. Plambeck, T.-C. Chang, T. Li, and G. Popping for useful discussions and thoughtful feedback. We gratefully acknowledge the James S. McDonnell Foundation, the National Science Foundation (NSF) and the University of Chicago for funding to construct the SZA. Partial support was provided by NSF Physics Frontier Center grant PHY-0114422 to the Kavli Institute of Cosmological Physics at the University of Chicago. Support for CARMA construction was derived from the states of California, Illinois, and Maryland, the James S. McDonnell Foundation, the Gordon and Betty Moore Foundation, the Kenneth T. and Eileen L. Norris Foundation, the University of Chicago, the Associates of the California Institute of Technology, and the National Science Foundation (NSF). Support for CARMA operations and analysis was provided in part by the National Science Foundation University Radio Observatories Program, including awards AST-1140019 (to the University of Chicago), AST-1140031 (University of California-Berkeley), AST-1140021 (California Institute of Technology), and by the CARMA partner universities.
\bibliography{bibliography}
\end{document}